\documentclass{article}

\usepackage{amsmath, amsthm, amssymb}
\makeatletter
\def\widebar{\accentset{{\cc@style\underline{\mskip10mu}}}}
\makeatother

\begin{document}
\numberwithin{equation}{section}
\def\bib#1{[{\ref{#1}}]}
\title{\bf A CFT description of the BTZ black hole: topology versus geometry (or thermodynamics versus statistical mechanics)}
\author{{ G.~Maiella ${ }^{1,2}$ and C.~Stornaiolo ${ }^{1,2}$}\\
{\em $~^{1}$ Istituto Nazionale di Fisica Nucleare,}
{\em Sezione di Napoli,}\\
 {\em  Complesso Universitario di Monte S. Angelo}
 {\em Edificio N' }\\ {\em via Cinthia, 45 -- 80126 Napoli}\\
{\em $~^{2}$ Dipartimento di Scienze
Fisiche,}\\ {\em Universit\`{a} ``Federico II'' di Napoli,}\\
 {\em  Complesso Universitario di Monte S. Angelo}
 {\em Edificio N'  }\\ {\em via Cinthia, 45 -- 80126  Napoli}\\ }

\maketitle

\begin{abstract}
In this paper we review the properties  of the black hole entropy in
the light of a general conformal field theory treatment. We find
that the properties of horizons of the BTZ black holes in ADS$_{3}$,
can be described in terms of an effective unitary CFT$_{2}$ with
central charge $c=1$ realized in terms of the Fubini-Veneziano
vertex operators.

It is found a relationship  between the topological properties of
the black hole solution and the infinite algebra extension  of the
conformal group in 2D, $SU(2,2)$, i.e. the Virasoro Algebra, and its
subgroup $SL(2,Z)$ which generates the modular symmetry. Such a
symmetry induces a duality for the black hole solution with angular
momentum $J\neq 0$. On the light of such a global symmetry we
reanalyze the Cardy formula for CFT$_{2}$ and its possible
generalization to $D>2$ proposed by E. Verlinde.
\end{abstract}

\vskip 0.5truecm
\section{Introduction}\label{introduction}
Gravity in 2+1 dimensions is a very  simple model where it is
possible to derive exactly the correspondence  between quantum black
holes properties and thermodynamical quantities.

In  presence of a negative cosmological constant
$\Lambda=-1/\ell^{2}$, the General Relativity in $2+1$ dimensions
admits the analogues of the $3+1$ Schwarzschild  and the Kerr black
hole known as the BTZ black hole \cite{btz1}.

In fact these solutions exhibit all the usual thermodynamic
properties of black holes \cite{Bekenstein:1972tm}. Their entropy is
found to be $1/4$ of
 the horizon area divided by $G _{3}$ the gravitational constant in
 three dimensions
\begin{equation}\label{horizon area}
    S=\frac{2\pi r_{+}}{4 G_{3}}.
\end{equation}

It satisfies the first law of thermodynamics, as it should,
\begin{equation}\label{firstlaw}
    dU= TdS+\Omega dJ
\end{equation}
where $U$ is identified with the ADM energy, $T$ is the Hawking
``temperature'', $J$ is the angular momentum and $\Omega$ is the
angular potential \cite{Carlip:2004mn}.

It is well known that for the $2+1$ gravity there are no bulk
dynamical degrees of freedom, or, in other words, all the dynamics
is in  the boundary. But there is a clear separation between the
gravitational sector and the matter sector.

In the following we analyze the black hole physical properties by
using the powerful tools of 2D conformal quantum field theory,
CFT$_{2}$; indeed, it is will be argued that the boundary degrees of
freedom of ADS$_{3}$ are described by an unitary representation of
the ``dual'' CFT$_{2}$, i.e. of the infinite Virasoro algebra
centrally extended with the central charge $c$ given by the
Brown-Henneaux relation \cite{Brown:1986nw}
\begin{equation}\label{centralcharge}
    c=\frac{3\ell}{2 G_{3}}.
\end{equation}

The paper is organized as follows. In sect. \ref{2} we discuss the
geometrical diffeomorphisms which preserve the conformal boundary
and its possible description by a Chern-Simons topological field
theory. The classical black hole general solution (which we shall
call in the following as  BTZ) is given in section \ref{3} together
with its properties. The explicit relation of the mass $M$ and the
angular momentum $J$ in terms of the generators of the classical
Virasoro algebra is found. In section \ref{4} the quantum version of
the BTZ solution in the Euclidean space is briefly analyzed.
Moreover it is argued that the relevant unitary representation of
CFT$_{2}$ is the $c=1$ Fubini-Veneziano vertex operator one
\cite{Fubini:1970xj}\ , following the Cardy argument
\cite{Cardy:1986ie}\ . In section \ref{5} we give a short summary of
the mathematical properties of the unitary representation of
CFT$_{2}$ which are used in this paper. In particular the Hilbert
space of the quantized ``momenta'' $\hat{p}$ and the winding
``numbers'' $\hat{w}$ for the Fubini-Veneziano scalar field
compactified on a circle $S_{1}$ is analyzed using  the SL(2,Z)
modular symmetry. That is reflected in a ``duality'' relation
between $(\ell, J)$ or $(r_{+},r_{-})$. In section \ref{6} we derive
the Brown-Henneaux relation by evaluating the quantum ``anomaly''
for the locally conformal (Weyl) transformations which is a
classical symmetry in both cases, for the gravitational equations in
the bulk and for CFT$_{2}$ defined on the boundary and identifying
the results \cite{Balasubramanian:1999re}\ . That is also a proof of
the validity of the ADS$_{3}$-CFT$_{2}$ duality guessed by Maldacena
\cite{Maldacena:1997re}.

The analogue of the Hawking temperature and entropy $S_{H}$ is
derived in section \ref{7} from the topology of spacetime,
identified to be $R^{2}\times S^{1}$. then as for the duality
between $(r_{+},r_{-})$, the topology of spacetime is strictly
related to the physical quantities of BTZ. In section \ref{8} we
stress how the modular invariance, SL(2,Z), of CFT$_{2}$, which
implies the Cardy equation (see sect. \ref{4}) is also crucial for
the validity of the cosmological Bekenstein bound
\cite{Bekenstein:1980jp} in 2D. We present a derivation of the
Casimir energy $E_{C}$ and entropy $S_{C}$ which should be relevant
for the possible extension to higher dimensions.

Some hints in such a direction are briefly discussed in section
\ref{9} where general comments and   conclusions are also given.

\section{The $\mathbf{2+1}$ anti-de Sitter spacetime
(ADS$_{3}$)}\label{2}

Before discussing the BTZ black hole, i.e. the black hole in 2+1
dimensions, let us introduce the ADS$_{3}$ spacetime. The reason is
that a black hole solution in 2+1 dimension, differently from the
Schwarzschild solution, has not a Newtonian asymptotic limit but an
ADS$_{3}$ one.

The ADS$_{3}$ is a vacuum solution of Einstein equations in 3
dimensions with a negative cosmological constant $\Lambda$.

Its metric in polar coordinates is
\begin{equation}\label{ads1}
     ds^{2}=\left[\left(\frac{r}{\ell}\right)^{2}+1\right]dt^{2}
     +\left[\left(\frac{r}{\ell}\right)^{2}+1\right]^{-1}dr^{2}+r^{2}d\varphi^{2}.
\end{equation}
where $\ell=|\Lambda|^{-1/2}$.

An ADS$_{3}$ is characterized by being a manifold with a (conformal)
boundary. In this case not all the diffeomorphisms are  allowed, but
only those which preserve the conformal boundary.

These can be identified with the conformal group SL(2,C), which is
the covering group of O(2,2) \cite{Ginsparg:1988ui}.

From the geometrical point of view, these diffeomorphisms are
generated by the vector fields

\begin{equation}\label{vector1}
    \xi^{(+)t}=\ell T^{+} +\frac{\ell^{3}}{2r^{2}}\partial^{2}_{u} T^{+}
    + O\left(\frac{1}{r^{4}}\right) \ \ \ \ \ \ \ \ \ \  \xi^{(-)t}=\ell T^{-} +\frac{\ell^{3}}{2r^{2}}\partial^{2}_{v} T^{-}
    + O\left(\frac{1}{r^{4}}\right)
\end{equation}

\begin{equation}\label{vector2}
    \xi^{(+)\varphi}=\ell T^{+} -\frac{\ell^{3}}{2r^{2}}\partial^{2}_{u} T^{+}
    + O\left(\frac{1}{r^{4}}\right) \ \ \ \ \ \ \ \ \ \  \xi^{(-)\varphi}=-\ell T^{-} +\frac{\ell^{3}}{2r^{2}}\partial^{2}_{v} T^{-}
    + O\left(\frac{1}{r^{4}}\right)
\end{equation}

\begin{equation}\label{vector3}
    \xi^{(+)r}=-r\partial_{u} T^{+}
    + O\left(\frac{1}{r}\right) \ \ \ \ \ \ \ \ \ \  \xi^{(-)r}=- r\partial_{v}
    T^{-}
    + O\left(\frac{1}{r}\right)
\end{equation}
where $T^{\pm}$ are functions of $u=t/\ell + \varphi$ and
$v=t/\ell-\varphi$.

The commutators $[\xi^{\pm}_{1},\xi^{\pm}_{2}]=\xi^{\pm}_{3}$ define
new vector fields of the form given above and  induce  two Virasoro
algebras with vanishing central charges on the functions $T^{\pm}$.

It is possible to show that the presence of a boundary modifies
these algebras by introducing a central charge $c$ related to $\ell$
as in eq. (\ref{centralcharge}).

Some properties of the ADS$_{3}$ spacetime are better described by
observing that the General Relativity action in $2+1$ dimensions is
equivalent to the Chern-Simons theory \cite{Carlip:2004mn}
\cite{Banados:1998sm}.

Then by doing  appropriate identifications the general relativistic
action with a negative cosmological constant  can be expressed  by
the following action

\begin{equation}\label{chernsimos}
  I_{CS}= \frac{k}{4 \pi}  \int_{M} Tr\left\{A\wedge dA+\frac{2}{3}A\wedge A \wedge A\right\}
\end{equation}
which can be split in the spatial and the time parts
\begin{equation}\label{chernsimossplit}
  I_{CS}= \frac{k}{4 \pi} \int dt \int_{\Sigma} Tr\left\{A\wedge  \dot{A}- A_{0}F \right\}
\end{equation}
where $A_{0}$ is a multiplier and the resulting field equation
\begin{equation}\label{fieldequation}
    F\equiv dA+A=0
\end{equation}
is a constraint. The solutions in the bulk take the trivial form
$A_{\mu}\sim G^{-1}\partial_{\mu} G$ and $\bar{A}_{\mu}\sim
(\partial_{\mu} \tilde{G}) \tilde{G}^{-1}$; where

\begin{equation}\label{G}
  G=   \left(%
\begin{array}{cc}
   \sqrt{r} & 0\\
  0 &  \frac{1}{\sqrt{r}} \\
\end{array}%
\right)g(u)
\end{equation}

\begin{equation}\label{G1}
 \tilde{G}=  \tilde{g}(v)  \left(%
\begin{array}{cc}
   \frac{1}{\sqrt{r}} & 0\\
  0 &  \sqrt{r} \\
\end{array}%
\right)
\end{equation}
are diffeomorphisms in the bulk and the connections are just trivial
gauge conditions.

But  these transformations are not anymore diffeomorphisms on the
boundary and the connections $A$ and $\bar{A}$,  are  the dynamical
variables of the theory. $g(u)$ and $\tilde{g}(v)$ are restricted
(see ref.) so that the ``affine'' algebra (Kac-Moody algebra)
currents $g^{-1}\partial_{\mu} g$ and $\partial_{\mu} g\cdot g^{-1}$
(i.e. the left and right currents of the algebra) induce the two
sectors of the Virasoro algebra, whose generators are

\begin{equation}\label{generatorsu}
   T_{uu}= \sum_{n}L_{n}
   e^{-inu}=\frac{k}{2}Tr(\sigma_{3}\partial_{u}A_{u}-A_{u}A_{u}),
\end{equation}

\begin{equation}\label{generatorsv}
   T_{vv}= \sum_{n}L_{n}
   e^{-inv}=\frac{k}{2}Tr(\sigma_{3}\partial_{v}\tilde{A}_{v}+\tilde{A}_{v}\tilde{A}_{v}).
\end{equation}

\section{Classical BTZ black hole}\label{3}
The black hole solution  in ADS$_{3}$ was found by Ba\~{n}ados,
Henneaux, Teitelboim and Zanelli in \cite{btz1}\cite{btz2}. The
corresponding quantum mechanical and thermodynamic properties are
discussed in \cite{Carlip:1994gc}.

In \cite{btz2} it is shown that an axially symmetric stationary
metric in a $2+1$ dimensional lorentzian spacetime (labelled with
the subscript Lor) with cosmological constant $\Lambda=-1/\ell^{2}$
is

    $$ds^{2}= -\left(
    -M_{Lor}+\frac{r^{2}}{\ell^{2}}+\frac{J_{Lor}^{2}}{4r^{2}}\right)dt^{2}
    +\left(-M_{Lor}+\frac{r^{2}}{\ell^{2}}+\frac{J_{Lor}^{2}}{4r^{2}}\right)^{-1}dr^{2}
    $$
    \begin{equation}\label{btzmetric1}
    +\,r^{2}\left(d\phi -\frac{J_{Lor}^{\phantom{2}} }{2r}dt\right)^{2}
\end{equation}
We can note the following features in this solution.

Differently from the $3+1$ solution the length scale is given by the
``curvature'' radius $\ell$, because the mass $M_{Lor}$ is a
dimensionless quantity.

The lapse function
\begin{equation}\label{lapse}
    N=-M_{Lor}+\frac{r^{2}}{\ell^{2}}+\frac{J_{Lor}^{2}}{4r^{2}}
\end{equation}
vanishes for
\begin{equation}\label{lapsezero}
     r_{\pm}^{2}= \frac{\ell^{2}M_{Lor}}{2}\left(
     1\pm\sqrt{1-\frac{J_{Lor}^{2}}{\ell^{2}M_{Lor}^{2}}}\right).
\end{equation}

It follows

\begin{equation}\label{massa}
     M_{Lor}=\frac{r_{+}^{2}+r^{2}_{-}}{\ell^{2}}
\end{equation}
and

\begin{equation}\label{momentoangolare}
     J_{Lor}=\frac{2r_{+}r_{-}}{\ell}.
\end{equation}

$g_{00}$  vanishes at
\begin{equation}\label{null}
     r=r_{erg}\equiv\ell M_{Lor}^{1/2},
\end{equation}
which  as for Kerr  solution called ergodic radius, which defines
the infinite red-shift surface of the black hole.

Finally at  $r=0$ there is a singularity on the causal structure but
not in the curvature, because the curvature is everywhere finite and
constant.

For large $r$ the BTZ metric (\ref{btzmetric1}) approaches the
ADS$_{3}$  metric (\ref{ads1}), then the asymptotic symmetry group
for this metric is  the conformal group in two dimensions $SO(2,2)$
or its covering $SL(2,C)$.

It is noteworthy that one can find some dual relations in the BTZ
black hole in presence of an angular momentum. Indeed the  second
degree algebraic equation  $N=0$ can be solved in terms of the
unknown $r^{2}$ or its of inverse $1/r^{2}$.  The solutions in terms
of $ r^{2}$ are related to the solutions of their inverse through
(\ref{momentoangolare}) according to which
\begin{equation}\label{dualr}
    r_{-}^{2}=\frac{J_{Lor}^{2}\ell^{2}}{4r_{+}^{2}},
\end{equation}
moreover in the  equation $N=0$  there is a dual relation  between
$J^{2}$ and $1/\ell^{2}$. This  relation, as of the well-known
duality relation in CFT$_{2}$, is a consequence of the $SL(2,Z)$
invariance of the unitary representations of Virasoro algebra in
CFT$_{2}$. We will discuss such fact in sect. 5.

The Lie algebra of the conformal group consists in two copies of the
Virasoro algebra with a central charge proportional to $\ell$, the
radius of curvature. It has generators $L_{n}$ and $ \bar{L}_{n}$
for any $n\in Z$. The two non zero charges for  metric
(\ref{btzmetric}) are
\begin{equation}\label{mass}
   M_{Lor}=\bar{L}_{0}+L_{0}
\end{equation}
and
\begin{equation}\label{angularmomentum}
    J_{Lor}=L_{0}-\bar{L}_{0},
\end{equation}
which are consistent with the fact that the eigenvalues of the
dilatation operators $L_{0}(\bar{L_{0}}$ are
$r^{2}_{+}/\ell^{2}(r^{2}_{+}/\ell^{2})$. An explicit derivation is
given in sect. \ref{5}. The ADS$_{3}$ is a maximally symmetric
space, it has $6$ Killing vectors. A spacetime with a black hole
shares with the AdS spacetime such symmetries only asymptotically,
but it generally has a lesser number of Killing vectors. Therefore a
black hole can be defined by its symmetries.

Given the Killing vectors $\mathbf{\xi}$, one can construct a
parameter subgroup such that to a given point $P$ we have $P\to
e^{t\mathbf{\xi}}P$. When $t$ is an integral multiple of a step
(conventionally one can take it as $2 \pi$) we define an
identification subgroup of $SO(2,2)$.

Correspondingly we can take the quotient space, which preserves the
properties of the  ADS$_{3}$. This quotient space is still a
solution of Einstein's equations.

If we label the coordinates by $x^{a}=(u,v,x,y)$, then the six
Killing vectors of the  ADS$_{3}$ are

\begin{equation}\label{killingvectors}
    J_{ab}=x_{b}\frac{\partial}{\partial x^{a}}-x_{a}\frac{\partial}{\partial x^{b}}
\end{equation}
In \cite{btz2} it is proved that the black hole solutions are
obtained by making the identifications defined previously by the
discrete group generated by the Killing vector
\begin{equation}\label{bhkilling}
   \mathbf{\xi}=\frac{1}{2}\omega^{ab}J_{ab}=\frac{r_{+}}{\ell}J_{12}-\frac{r_{-}}{\ell}J_{03}-J_{13}
   +J_{23}.
\end{equation}
where $\omega^{ab}$ is an antisymmetric tensor, defined by the
preceding relation, with eigenvalues $\pm r_{+}/\ell$, and $\pm
r_{-}/\ell$.

The Casimir invariants are
\begin{eqnarray}
  I_{1} &=&  \omega_{ab}\omega^{ab}=-\frac{2}{\ell^{2}}(r^{2}_{+}+r^{2}_{-})=-2M \\
  I_{2} &=&  \frac{1}{2}\epsilon_{abcd}\omega^{ab}\omega^{cd}=-\frac{4}{\ell^{2}}
  =-2\frac{|J|}{\ell}.
\end{eqnarray}
which are respectively proportional to the quantities defined in
(\ref{mass}) and (\ref{angularmomentum}).
\section{ BTZ black hole quantum mechanics: from Minkowski to the Euclidean
description}\label{4}

Sometimes it is more convenient to work in an Euclidean frame, where
 the Euclidean ``time'' is  $t\rightarrow \tau=it$, and we put $M_{Lor}=M$ and
$J_{Lor}=iJ$

   $$ ds^{2}= -\left(
    -M +\frac{r^{2}}{\ell^{2}}-\frac{J ^{2}}{4r^{2}}\right)d\tau^{2} +\left(
    -M +\frac{r^{2}}{\ell^{2}}-\frac{J ^{2}}{4r^{2}}\right)^{-1}dr^{2}$$
    \begin{equation}\label{btzmetric}+\,
    r^{2}\left(d\phi -\frac{J}{2r}d\tau\right)^{2}.
\end{equation}

The singularities of this metric are in $r=0$, $r=\infty$ and in
\begin{equation}\label{singularities}
     r_{\pm}^{2}= \frac{\ell^{2}M}{2}\left(
     1\pm\sqrt{1+\frac{J^{2}}{\ell^{2}M^{2}}}\right).
\end{equation}
In the limit  $J\to 0$, the metric reduces to
\begin{equation}\label{btzmetricjzero}ds^{2}= \left(
    -M +\frac{r^{2}}{\ell^{2}}\right)d\tau^{2} +\left(
    -M +\frac{r^{2}}{\ell^{2}}\right)^{-1}dr^{2}
    +\,r^{2}d\phi^{2}.
\end{equation}
 the metric is not  singular at $r=0$  and the singularities
(\ref{singularities}) reduce to $r_{+}=\sqrt{M}\ell$ and $ r_{-}=0$,
which is the equivalent in $2+1$ dimensions of the Schwarzschild
radius in $3+1$ dimensions \cite{btz2}.

By appropriate changes of coordinates we can put in evidence the
properties of the metric (\ref{btzmetricjzero}) in this limiting
case,
%
it is possible to show the existing of  the periodicity condition
$\theta\sim \theta+2\pi$ and
 the identifications of the coordinates lead to quotient the upper semispace
 by identifying $R\sim e^{2\pi\sqrt{M}} R$, one can make a further
 change of metric,and work in the region delimited between the upper hemisphere
 $R=1$ and $R=e^{2\pi\sqrt{M}}$, where the singular points are
 identified through the radial lines, the manifold so defined is a solid
 torus. The extension of this procedure to the case $\mathbf{J}\neq
0$ is straightforward.

Going back to the CFT$_{2}$ for the boundary dynamics, we stress a
very relevant result. It has been proven in \cite{Cardy:1986ie} that
the density of states is given by
\begin{equation}\label{density}
    S=2\pi \left[ \left(\frac{c L_{0}}{6}\right)^{1/2}+\left(\frac{c
    \bar{L}_{0}}{6}\right)^{1/2}\right].
\end{equation}

In \cite{Martinec:1998wm} the Cardy equation was used to derive an
effective CFT$_{2}$ with $c=1$ described by the Coulomb gas (vertex)
operator (for details see section \ref{5}) as a consequence of
modular invariance, $SL(2,Z)$, and other very general properties of
CFT$_{2}$. Then  the ``highest weight states'' of such  CFT$_{2}$
can be considered as the microstates of the gravity in $2+1$
dimensions, while the global states are described by the boundary
dynamics. More precisely the gravity is determined only by the
global geometric data and does not have ``local excitations'',
however  eq. (\ref{density}) sets for ADS$_{3}$  spacetime the
correspondence with local states of the (conformal) field theory
which are the ``microscopic'' excitations for $2+1$ gravity.

This point of view resembles the one advocated by Martinec in
\cite{Martinec:1998wm} and it will be made precise in the following
section.

\section{Short summary of unitary representations of CFT$\ _{2}$}\label{5}

To analyze the properties of the unitary representations of
CFT$_{2}$, it is customary to use the Euclidean spacetime.

In complex coordinates the metric is
\begin{equation}\label{complexcoordinates}
     ds_{E}^{2}=dzd\bar{z}
\end{equation}
where $z=x+iy$ and $\bar{z}=x-iy$. Being the field theory conformal
invariant one can split all the field in an analytic and an
antianalytic part, i.e. for $\Phi(x,y)$ one can write
\begin{equation}\label{split}
\Phi(x,y)=\Phi_{L}(z) + \bar{\Phi}_{R}(\bar{z}).
\end{equation}
Henceforth  we discuss only the analytic part, i.e. the left sector.

Relevant conformal fields called highest weight states are
\begin{description}
  \item[a\ )] The energy-momentum tensor $T_{ab}(x,y)$ which is an
  operator of conformal dimension $2$ and it is written as
  \begin{equation}\label{emtensorconformal}
    T(x,y)= T_{zz} (z) + \bar{T}_{\bar{z}\bar{z}}
    (\bar{z})=T_{L}(z)+\bar{T}_{R}(\bar{z}).
  \end{equation}

  \item[b\ )]  The currents  $J_{a}(z)$ which have conformal dimension $1$
  and are generators of symmetries.
\end{description}

They are a necessary ingredient for the solvability of CFT$_{2}$,
for details see \cite{Ginsparg:1988ui} and
\cite{DiFrancesco:1997nk}.

Then one can define the Virasoro algebra for the left sector as
follows

\begin{equation}\label{Virasoroalgebracft}
    [L_{n},L_{m}]=(n-m)L_{n+m}+ cn(n^{2}-1)\delta_{n+m;0}
\end{equation}
where
\begin{equation}\label{elleenne}
    L_{n}=\frac{1}{2\pi i}\oint z^{n}T_{L}(z).
\end{equation}
Naturally $\bar{L}_{n}$ defined for the antianalytic part defines
the same algebra.

 Notice that the central extension, i.e. the second term in eq.
 (\ref{Virasoroalgebracft}) is absent for $n=0,\pm 1$ as it should
 be. In fact $L_{0},\, L_{\pm1}$ generate the 2-D conformal group
 $SO(2,2)$ (whose covering group is $SL(2,C)$). But the relevance of
the entire algebra is a milestone of theoretical physics (see
\cite{Ginsparg:1988ui}).

The Operator Product Expansion (OPE) is another crucial technique of
conformal field theory in any dimension. For example from simple
scaling arguments one derives the
\begin{equation}\label{tz}
    T(z)J(w)\mathop{\rightarrow}\limits_{(z-w)\to
    0}\frac{J(w)}{(z-w)^{2}}+\frac{\partial_{w}J}{(z-w)}+\rm{
    regular\ \
    terms}
\end{equation}

A $CFT_{2}$ is completely and exactly known if one can derive all
the highest weight states and the associated operators $O_{a}(z)$
$a=1,\dots k$ and their 3 point functions exactly
\begin{equation}\label{operators}
    O_{a}(z)O_{b}(w)=\lim_{z\to w}\frac{O_{c}(w)}{(z-w)^{d_{a}+d_{b}+d_{c}}}
\end{equation}
where $d_{a}$ are the conformal dimensions.

We refer for the complete analysis of such beautiful results to
\cite{Ginsparg:1988ui} \cite{Cristofano:1993nb}.

For our purpose we recall only the results concerning the
representation of the $c=1$ $CFT_{2}$ .

The highest weight states can be described by the vertex operators
\cite{Fubini:1970xj},

$$ V(z)=:e^{i\alpha\Phi(z)}:$$

We will see that for rational CFT$\ _{2}$ the values of $\alpha^{2}$
are rational numbers. The scalar real field \cite{Fubini:1970xj} is
represented in Fock Space as follows
\begin{equation}\label{fub}
   \Phi(z)= a_{0}+\rho_{0}ln z+\sum_{n=-\infty; n\neq 0}^{\infty}
   a_{n}z^{n}
\end{equation}
where $a^{\dag}_{n}=a_{-n}$ due to the reality of $\Phi$. The
commutator relations are given by
\begin{equation}\label{crel}
    [x_{0},p_{0}]=1;\ \ \ \  [a_{n}, a_{-m}]=\delta_{n+m;0}
\end{equation}
The Green function $G(z)$ is evaluated as
\begin{equation}\label{greenfunct}
    [\Phi(z),\Phi(w)]=ln \frac{(z-w)}{u}
\end{equation}
$u$ being an infrared cut-off.

A very interesting case is when the scalar field $\Phi(z)$ is
compactified on a circle $S_{1}$ with radius $R$ Then the highest
weight states are defined by the Hilbert space as
\begin{equation}\label{hws}
    \hat{p}\ |l>=\frac{l}{R}|l>
\end{equation}
and for the ``dual'' state by the winding number $\hat{w}$
\begin{equation}\label{dualstate}
 \hat{w}|k>=k R|k>
 \end{equation}
So we get a $U(1)$ symmetry for the highest weight states enhanced
to $U(1)\times U(1)$. In fact it can be shown that the conformal
weights $\Delta$ and $\bar{\Delta}$ for the $R$, $L$ sector are
\begin{equation}\label{sum}
\Delta+\bar{\Delta}=\frac{\hat{p}^{2}}{R^{2}}+R^{2}\hat{w}^{2}
\end{equation}
\begin{equation}\label{subtract}
\Delta-\bar{\Delta}= 2\hat{p}\cdot \hat{w}\, .
\end{equation}
from which
\begin{equation}\label{pesiconformi}
    \Delta= \left(\frac{\hat{p}}{R}+R\hat{w} \right)^{2}\; ;\ \ \ \ \ \bar{\Delta}=\left(\frac{\hat{p}}{R}-R\hat{w} \right)^{2}
\end{equation}

If $R^{2}=m$ and $m\ \in Z_{+}$  then $(l,k)\leq m$. In other terms
the highest weight states are finite and all the correlation
functions are given by products of binomials as $(z-w)^{\alpha^{2}}$
with $\alpha^{2}$ a positive integer. Furthermore $\hat{p}$ and
$\hat{w}$ can be interpreted as charged in the so-called 2D Coulomb
gas interpretation of the vertex operators \cite{DiFrancesco:1997nk}
\cite{Cristofano:1992yy}. Eqs. (\ref{sum}) and (\ref{subtract}) for
$R^{2}=1$ reproduce respectively eqs. (\ref{massa}) and
(\ref{momentoangolare}) if we identify
\begin{equation}\label{pew}
\hat{p}=\frac{r_{+}}{\ell}\; ; \ \ \ \ \ \ \ \
\hat{w}=\frac{r_{-}}{\ell}
\end{equation}

from which
\begin{equation}\label{deltapiudelta}
  \Delta+\bar{\Delta}= \hat{p}^{2}
  +\hat{w}^{2}=\frac{r^{2}_{+}+r^{2}_{-}}{\ell^{2}}=M
\end{equation}
\begin{equation}\label{deltamenodelta}
    \Delta-\bar{\Delta}= 2\hat{p}\cdot
\hat{w}=2r_{+}r_{-}=J\ell\, .
\end{equation}

then
\begin{equation}\label{ddelta}
    \Delta= \left(\hat{p}+\hat{w}\right)^{2}= \frac{(r_{+}+r_{-})^{2}}{\ell^{2}}
\end{equation}
and
\begin{equation}\label{ddeltabarra}
   \bar{ \Delta}= \left(\hat{p}-\hat{w}\right)^{2}= \frac{(r_{+}-r_{-})^{2}}{\ell^{2}}
\end{equation}

 For $J=0\Leftrightarrow r_{-}=0$ one gets $\Delta=\bar{\Delta}$. In
 the particular case of maximal $J$ i.e. $r_{+}=r_{-} $ one finds
\begin{equation}\label{xxx}
    \Delta= \frac{4}{\ell^{2}} r_{+}^{2}\, , \ \ \ \ \ \ \ \ \ \ \
    \bar{\Delta} =0.
\end{equation}
We notice that for such extremal case the CFT$_{2}$ contains  one
sector or in other terms it is a chiral theory. In such a case it
looks similar to the one used in \cite{Cristofano:1992yy} to
describe the Laughlin anyons  for a Quantum Hall fluid
\cite{Cristofano:1990fg}. Such a connection has previoulsy been
noticed, in a different context, in \cite{Balachandran:1994up}.

\section{Cosmological constant and central charge}\label{6}

As it is well known there have been different ways to analyze the
gravitational properties of the anti de Sitter space in $2+1$
dimensions.

Here we shall start from a very interesting fact derived in the work
by Brown  and Henneaux, i.e. the relation between the cosmological
constant $\Lambda=- 1/ \ell^{2}$ in $AdS_{3}$ and the central charge
of the boundary $CFT_{2}$, equation (\ref{centralcharge}). We shall
briefly summarize the derivation as given in
\cite{Balasubramanian:1999re}. In fact starting from the action
\begin{equation}\label{action}
    S= -\frac{1}{16 \pi G} \int_{M} d^{3} x \sqrt{g} \left[R-\frac{d(d-1)}{\ell^{2}}
    \right]-\frac{1}{8 \pi G}\int_{\partial M} \sqrt{-\gamma}\Theta
    +\frac{1}{8 \pi G} S_{ct}(\gamma_{\mu\nu})
\end{equation}
where $\Theta$ is the trace of the extrinsic curvature of the
boundary. $M$ is  $AdS$ and $\partial M$ is its boundary. From
(\ref{action}) one gets
\begin{equation}\label{emtensor1}
    T^{\mu\nu}=\frac{1}{8 \pi G}\left[ \Theta^{\mu\nu}-\Theta\gamma^{\mu\nu} +
    \frac{2}{\sqrt{-\gamma}}\frac{\delta S_{ct}}{\delta\gamma_{\mu\nu}}\right]
\end{equation}
where  $S_{ct}$ has the role of canceling the divergences when
$\delta\bar{M}$ goes to the $AdS $ boundary $\delta M$.

By a careful analysis one finds
\begin{equation}\label{emtensor2}
    T^{\mu\nu}=\frac{1}{8 \pi G}\left[ \Theta^{\mu\nu}-\Theta\gamma^{\mu\nu}  -
    \frac{1}{\ell}\gamma^{\mu\nu} \right]
\end{equation}
where all the quantities refer to the boundary metric and
\begin{equation}\label{einst}
    G_{\mu\nu}= R_{\mu\nu}-\frac{1}{2}R g_{\mu\nu}.
\end{equation}
Remind that in $2d$  $R_{\mu\nu}=g_{\mu\nu} R$. For $AdS_{3}$ one
can compute the mass and angular momentum of  BTZ black holes.

Let us introduce the $AdS_{3}$ metric
\begin{equation}\label{ads3}
    ds^{2}=\left(\frac{\ell}{r}\right)^{2}dr^{2}+\left(\frac{r}{\ell}\right)^{2}(dx^{2}-dt^{2})
    .
\end{equation}
At a fixed $r$ we have a boundary conformal to $R^{1,1}$ i.e.
\begin{equation}\label{bound}
    -g_{tt}=g_{xx}=\left(\frac{r^{2}}{\ell^{2}}\right).
\end{equation}
Following such line of thought one reproduces general results
usually derived with conventional technique see ref [] for
details.

In particular when $M=-1/8 \pi G$; $J=0$ the $BTZ$ metric approaches
the global $AdS_{3}$, while $M=0$ and $\mathbf{J}=0$ it is similare
to the Poincar\'{e} metric.

Let us derive the Weyl anomaly, in a covariant way. As it is
well-known for Euclidean  $CFT_{2}$ with metric is $ds^2=dzd\bar{z}$
the diffeomorphisms are defined as by
\begin{eqnarray}\label{reparam}
 \nonumber
  z &\to & z-f(z) \\
  \bar{z} &\to &  \bar{z}-g(\bar{z})
\end{eqnarray}
and
\begin{eqnarray}\label{eqnarray}
 \nonumber  T_{zz} &\to&  T_{zz} +(2\partial_{z} f(z)  T_{zz} + z \partial_{z}
  T_{zz}-\frac{c}{24 \pi}\partial^{+}_{z}f\\
   T_{\bar{z}\bar{z}} &\to & T_{\bar{z}\bar{z}} +(2\partial_{\bar{z}} g(\bar{z})
   T_{\bar{z}\bar{z}} + \bar{z} \partial_{\bar{z}}
  T_{\bar{z}\bar{z}}-\frac{c}{24 \pi}\partial^{+}_{\bar{z}}g
\end{eqnarray}
where $z=x+iy$ and $\bar{z}=x-iy$.

Really equation (\ref{reparam}) is a symmetry of the $CFT_{2}$ only
at a classical level. At the quantum level. In fact we have to
introduce an ultraviolet cut-off $u$ in quantum time-like convergent
loops and show that a field theory in ADS$_{3}$ remains invariant if
we rescale $z\to z'=e^{\lambda}z$ ($\lambda>0$). Equivalently the
metric should be Weyl rescaled to preserve $ds^{2}=-dzd \bar{z}$.
Starting from eq. (\ref{ads3}), if we consider the diffeomorphism
eq. (\ref{reparam}) there is a Weyl scaling of the boundary metric

Then we require that the asymptotic form  (for $r^{2}\to \infty$)
remains invariant. One can prove that it is so if for $r^{2}\to
\infty$
\begin{eqnarray}\label{metricacomponenti}
    g_{zz}=-\frac{r^{2}}{2}\ \ \ \ \ \ \ \  g_{++}=g_{--}= O(1)\\
     g_{rr}=\frac{\ell^{2}}{r^{2}}+O\left(\frac{1}{r^{4}}\right)
     \ \ \ \ \ \ \ \ g_{+\,r}=g_{-\,r}=O\left(\frac{1}{r^{3}}\right)
\end{eqnarray}

From now on the boundary $CFT_{2}$ is analyzed in the Euclidean
metric as usually one does for the relativistic quantum field
theory.

With these diffeomorphisms  the metric changes as
\begin{equation}\label{metricchanges}
    ds^{2}\to
    \frac{\ell^{2}}{r^{2}}dr^{2}-r^{2}dzd\bar{z}-\frac{\ell^{2}}{2}
    (\partial^{3}_{z}\xi^{+})dz^{2}
    -\frac{\ell^{2}}{2}(\partial^{3}_{\bar{z}}\xi^{+})d\bar{z}^{2}
\end{equation}
and we can compute the stress-energy tensor which is
\begin{equation}\label{stressenergytensor}
    T_{zz}=-\frac{\ell}{16 \pi G}\partial^{3}_{z}\xi^{z} \ \ \ \ \
    \ \ \ \ \ T_{\bar{z}\bar{z}}=-\frac{\ell}{16 \pi
    G}\partial^{3}_{\bar{z}}\xi^{\bar{z}}
\end{equation}

Equations (\ref{eqnarray}) do agree with (\ref{stressenergytensor})
if
\begin{equation}\label{centralcharge1}
   c=\frac{3 \ell}{ 2 G}
\end{equation}
according to (\ref{centralcharge}).

Naturally the analysis above is in agreement with the
ADS$_{3}$/CFT$_{2}$ duality on which we have not much to say here.
We stress  that  $r $ plays the r\^ole of the ultraviolet cut-off in
the general relativity analysis of the BTZ metric as   the usual
cut-off $a$  does on the quantum field theory (CFT$_{2}$) side.

\section{Thermodynamics and topology of a BTZ black hole } \label{7}

We will show that the BTZ black hole is a thermodynamic object with
``effective'' temperature
\begin{equation}\label{Temper}
     T_{0}=\frac{1}{\beta_{0}}=\frac{r_{+}^{2}-r_{-}^{2}}{2\pi \ell
     r_{+}} .
\end{equation}
as one would guess by analogy to the Schwarzschild case.

To do so we can apply the Euclidean path integral method
\cite{Gibbons:1994cg}. In complete (and straightforward) analogy to
the $3+1$ dimensional case one finds  the Euclidean metric
(\ref{btzmetric}) (with $t=i\tau$) introduced in section \ref{3}.
Such metric is singular at
\begin{equation}\label{root1}
    r^{2}_{+}=\left\{\frac{M\ell^{2}}{2}\left[1+\left(1+\frac{J_{E}^{2}}{M\ell^{2}}\right)^{1/2}\right]\right\}
\end{equation}
and
\begin{equation}\label{root2}
    r^{2}_{-}\equiv[-i|r_{-}|]^{2}=\left\{\frac{M\ell^{2}}{2}
    \left[1-\left(1+\frac{J_{E}^{2}}{M\ell^{2}}\right)^{1/2}\right]\right\}.
\end{equation}

As shown in \cite{Carlip:1994gc} the metric (\ref{btzmetric}) is
positive definite of constant negative curvature; then  it is
isometric to the hyperbolic three-space $H^{3}$. With a coordinate
change one obtains the metric of the standard half-space of $H^{3}$,
i.e.

$$ds^{2}= \frac{\ell^{2}}{z^{2}}(dx^{2}+dy^{2}+dz^{2}),\ \ \ \ \
     z>0$$
\begin{equation}\label{cartesian1}
 \phantom{ds^{2}abcd }\ \   \ \ \ \ = \frac{\ell^{2}}{\sin^{2}\chi}\left(\frac{dR^{2}}{R^{2}}
  +d\chi^{2}+cos^{2}\chi d\vartheta^{2} \right);\ \ \    z>0
\end{equation}
with the identifications
\begin{equation}\label{identifi}
    (R,\vartheta,\chi)\sim (Re^{2\pi r_{+}/\ell},\vartheta +2\pi
    |r_{-}|/\ell, \chi),
\end{equation}
the resulting topology is $R^{2}\times S^{1}$ as expected.

The requirements of smoothness lead to
\begin{equation}\label{smoothness}
    (\Phi,\tau)\sim (\Phi+ m \tilde{\Phi}, \tau+ \beta_{0})
\end{equation}
    \begin{eqnarray}
       \tilde{\Phi}=\frac{2\pi r_{-}\ell}{r^{2}_{+}-r^{2}-} & ,&\beta_{0}= \frac{2\pi r_{+}\ell^{2}}{r^{2}_{+}-r^{2}_{-}},  \\
        \tilde{\Phi}(r_{-}=0)=0& ,&\beta_{0}(r_{-}=0)=
        \frac{2\pi\ell^{2}}{r_{+}}.
    \end{eqnarray}
For a temperature $T_{0}=\beta_{0}^{-1}$ and a rotational chemical
potential $\Omega$ then
\begin{equation}\label{boh}
    I_{E} = 4\pi r_{+} - \beta_{0}(M-\Omega J)
\end{equation}
then one has
\begin{equation}\label{entropy}
    S\equiv S_{E}=\frac{2\pi r_{+}}{4\ell}
\end{equation}
which is the analogous of the Bekenstein-Hawking entropy, in $2+1$
dimensions. Therefore we can identify $T_{0}$ with $T_{BH}$ and $S$
with $S_{BH}$.

\section{Cardy formula and the Verlinde proposal}\label{8}
We have seen in sect. 4 that the entropy of a  CFT$_{2}$ (the Cardy
formula) depends crucially on the central charge $c$. We shall write
it in a simplified way as
\begin{equation}\label{cardynew}
   S=2\pi \sqrt{\frac{c}{6}\left(L_{0}-\frac{c}{12}\right)}.
\end{equation}
The term $c/12$ is the vacuum energy of the system.

Moreover the central charge $c$, for a system of finite volume, is
strictly related to the boundary (surface) energy, i.e. the Casimir
energy $E_{c}$. Such a quantity is not an extensive term, i.e.
proportional to the volume $V$, but a subextensive one.

All the previous statements have been proved to be true for
 a large class of CFT$_{2}$ \cite{Cardy:1986ie}. Finally it has been
argued recently by Verlinde \cite{verlinde} that eq.
(\ref{cardynew}) can be generalized for any dimension $D$  if the
central charge $c/12$ is replaced by the Casimir energy [see also
ref 3].

The main support of such an assumption consists in  relating the
Cardy formula to the cosmological Beckenstein bound
\cite{Bekenstein:1980jp}
\begin{equation}\label{beckbound}
    S\leq S_{BH}.
\end{equation}
Here $S$ is the entropy of the entire system, while  $S_{BH}$ is the
Beckenstein entropy, which for any $D$ can be defined as
\begin{equation}\label{beckensteinentropy}
    S_{BH}=\frac{2\pi}{D-1}E\,R
\end{equation}
Such a bound is quite restrictive for low energy density, see
\cite{Kraus:2006wn}. To clarify the origin of equation
(\ref{beckensteinentropy}) we can apply a  simple scaling argument
\cite{verlinde} by observing that , for $V_{D}\rightarrow \lambda
V_{D}$, $E\rightarrow \lambda E$ and $S_{BH}\to
\lambda^{1+\frac{2}{d}}S_{BH}$.

By looking more closely to the role of Casimir energy in the
cosmological bounds (see \cite{verlinde} and \cite{Savonije} for
details) we will see that eq. (\ref{beckbound}) has a very
surprising physical interpretation.

To this aim we shall repeat a general argument. The entropy of a
thermodynamical system $S_{V}$ and the associated energy $E_{V}$ are
extensive quantities, i.e. proportional to the volume $V$ in $D$
dimensions.

That simply implies the relation
\begin{equation}\label{thermo}
   \rho+p=Ts.
\end{equation}
where $\rho$, $p$ and $s$ are respectively the energy density, the
pressure and the entropy density.

 Now the extensiveness of $E$ means that  $E(\lambda V, \lambda
E)=\lambda E(S,V)$. By differentiating one finds
\begin{equation}\label{extensi}
   E=V\left( \frac{\partial E}{\partial V}\right)_{S}+S\left( \frac{\partial E}{\partial S}\right)_{V}
\end{equation}

for  $\lambda=1$.

The first law of thermodynamics tells that
\begin{equation}\label{first1}
\left( \frac{\partial E}{\partial V}\right)_{S}=-p
\end{equation}
i.e.
\begin{equation}\label{firstlaw}
    TS=E+pV.
\end{equation}
But for a system with a boundary  there is a surface energy, the
Casimir energy, which implies a non-extensive contribution, $S_{C}$,
to tne entropy $S$, as it was already stated.

At this point we stress that $S_{C}$ is a finite size effect due to
quantum fluctuations, then its contribution is proportional to the
area $A$ of the boundary is not zero at any temperature, included
$T=0$. It is appropriate to notice here the strong resemblance with
the physics of quantum phase transitions recently so much studied
\cite{qpt} and \cite{Vidal:2002rm}. The origin and physical
relevance of the area law and/or of the Black Hole entropy seems
very surprising in these papers.

This fact is crucial for understanding the relation between the
topology of a $2+1$ dimensional spacetime and the thermodynamics of
the BTZ black hole. Now we will analyze how to generalize some of
the previous results for $D>2$.

A possible definition of Casimir energy in any spatial dimension $D$
can be the following
\begin{equation}\label{casimirenergydefinition}
    E_{C}=(D-1)(E+pV-TS)
\end{equation}
i.e the term which violates the Euler identity eq. (\ref{extensi}).
This definition is quite useful in our context.

A simple dimensional analysis tells us that $E_{C}$ scales as
\begin{equation}\label{scalingofEC}
   E_{C}(\lambda S, \lambda V)=\lambda^{1-2/(D-1)}E_{C}(S,V)
\end{equation}
Notice that the exponent of $\lambda$ depends on the dimensions $D$
in contrast to the extensive quantities.

Finally we can separate the two terms in $E$
\begin{equation}\label{etotale}
    E=E_{ext.}+\frac{1}{2}E_{C}
\end{equation}
where the extensive term is $E_{ext.}$ and the non extensive one is
$E_{C}$, the factor $1/2$ is introduced for convenience.

Now we come back to the bold assumption about $S_{c}$ and the
definition of $S_{B}$ in eq. (\ref{beckensteinentropy}). We notice
that $ER$ is independent of $V$ as a consequence of conformal
invariance. Therefore it is only a function of $S$.  Naturally both
terms $E_{ext.}$ and $E_{C}$ are only functions of $S$, more
precisely by usual scaling arguments one can write
\begin{equation}\label{scalingargum}
    E_{ext.}=\frac{a}{4\pi R}S^{1+1/(D-1)}\; ; \ \ \ \ \ \ \ \  E_{C}=\frac{b}{2\pi
    R}S^{1-1/(D-1)}
\end{equation}
with $(a,b)$ positive constants.

From that the entropy $S$ is given by
\begin{equation}\label{entropys}
    S=\frac{2\pi R}{\sqrt{ab}}\sqrt{E_{C}(2E-E_{C})}=\frac{2\pi R}{\sqrt{ab}}\sqrt{2E_{C}E_{ext.}}.
\end{equation}
That is exactly the Cardy formula (up to normalization) if one makes
the identifications
$$\mathrm{1)}\, \ \ L_{0} \to  ER  $$
 \begin{equation}\label{substitutions}
     \;\mathrm{2)}\,\ \   \frac{c}{6}  \to E_{C}R
 \end{equation}

Notice that we have used only conformal invariance and scaling
arguments  to derive eq. (\ref{entropys}) then it is very tempting
to assume that it is  true for any $D>2$.

For fixed $E$ eq. (\ref{entropys}) has a maximum when $E=E_{C}$,
i.e.
\begin{equation}\label{maximum}
    S\leq \frac{2\pi}{\sqrt{ab}}ER.
\end{equation}
That is the Beckenstein bound up to a constant term.

Many of our results are nice exemplifications of Maldacena
ADS$_{k}$/CFT$_{k-1}$ correspondence for ADS$_{3}$. More precisely
the thermodynamics of CFT$_{2}$ is identified with the
thermodynamics of the BTZ black hole as argued in
\cite{Witten:1998zw}.

\section{Comments and conclusions}\label{9}
In this paper we have emphasized the role and the strict relations
between the different properties,

1) the origin of the Brown-Henneaux relation (\ref{centralcharge})
is found to be the Weyl (trace) anomaly which for CFT$_{2}$ is
proportional to the central charge of the Virasoro algebra (sect.
6).

A related (dual) derivation of the Weyl anomaly
(\ref{metricacomponenti}) gives a contribution proportional to
$\ell/G$. The eq. (\ref{centralcharge}) relates the (quantum) vacuum
energy in CFT$_{2}$ with the (classical) gravity vacuum energy in
ADS$_{3}$ parameterized by $\ell$.

2) The metric of the space-time  ADS$_{3}$and its symmetries (i.e.
the geometry of space-time) implies the symmetries (diffeomorphisms)
i.e. the classical conformal group $SO(2,2)$ at the boundary

3) The quantum extension of this symmetry, the Virasoro algebra, is
assumed to be true for the boundary of ADS$_{3}$ ( assumption needed
in order to reproduce the Weyl anomaly).

Therefore the role of the global symmetry $SL(2,Z)$ (the modular
invariance of CFT$_{2}$) is to generate a ``duality'' relation
between $r_{+}$ and $r_{-}$, eq. (\ref{pew}) or equivalently between
$\ell$ and $J$ very reminiscent of the electro-magnetic duality in
the description of the Quantum Hall effect by CFT$_{2}$(see ref.
\cite{Cristofano:1992yy}). This fact needs a deeper understanding.

4) The thermodynamics of  CFT$_{2}$ now gives the results of sect. 7
where the Bekenstein-Hawking temperature $T_{BH}$, and the related
``entropy'' eq. (\ref{beckensteinentropy}) are evaluated to be the
correct ones.

Surprisingly those results are consequence of the topology of the
space-time metric when a  BTZ black hole of mass $M$ and angular
momentum $J$ is present. Then topology implies thermodynamical
properties of CFT$_{2}$ and viceversa. Moreover the black-hole
physical quantities eqs.  (\ref{massa}) and (\ref{momentoangolare})
are all expressed in terms of $\ell$ and become zero when $c\to 0$ (
$E_{C}\to 0$) or equivalently when $\ell\to 0$. That seems to be a
further support for the validity of the Cardy formula for $D>2$ or
of a related more general formula.

These results emphasize the relevance of the Casimir energy $E_{C}$
for the cosmological implication of the Cardy formula eq.
(\ref{cardynew}) analyzed in papers  \cite{verlinde} and
\cite{Savonije}. Our result shows that in 2D the ADS$_{3}$ entropy
$S\equiv S_{BH}$ as derived in section \ref{7} does saturate the
Bekenstein bound (\ref{beckbound}) implying that $E=E_{C}$ which, of
course, is the maximum value of the Casimir energy.

Then for $D>2$ one should find black hole solutions which do not
saturate the bound. Therefore it seems quinte important to deepen
the study of black hole phase transitions as the Hawking-Page
\cite{Hawking:1982dh}, started in \cite{Savonije}
\cite{Kutasov:2000td} and \cite{Cappiello:2001tf}, to understand the
evolution of its physical quantities as $( r_{+},r_{-})$ or $(M,J)$
as functions of $E_{c}$ or $\ell$.

It is our opinion that there is a strong interplay between the
physics of Quantum Phase Transitions, the Hawking-Page phase
transition and the thermodynamics of Black Holes as one can infer
form the work done by the community of cosmologists
\cite{Bombelli:1986rw}\cite{Bekenstein:1994bc}, the community of
statistical mechanics and field theorists
\cite{qpt}\cite{Vidal:2002rm}. In this contest one of the main
problems is to clarify the relation between the various definitions
of entropy as pointed out in \cite{Bekenstein:1994bc}. The problem
of ``information loss'' can be also carefully analyzed in this
framework. There are already clear indications in favor of the
recent work by Hawking \cite{Hawking:2005kf} in which the unitarity
of the black hole theory is preserved. See also the paper of
\cite{'tHooft:1993gx}.

\end{document}